\documentclass{PoS}

\usepackage{amsmath}

\newcommand{\bmat}{\begin{pmatrix}}
\newcommand{\emat}{\end{pmatrix}}

\title{Correlations and textures in the neutrino mass matrix}

\ShortTitle{Correlations and textures in the neutrino mass matrix}

\author{Walter Grimus and \speaker{Patrick Otto Ludl}%
        \\
        University of Vienna, Faculty of Physics, Boltzmanngasse 5, A--1090 Vienna, Austria\\
        E-mail: \email{walter.grimus@univie.ac.at}, \email{patrick.ludl@univie.ac.at}}

\abstract{The recent enormous improvement of our knowledge of the
neutrino oscillation parameters
has motivated us to reinvestigate the allowed ranges
of the elements of the neutrino mass matrix in the basis where the
charged-lepton mass mass matrix is diagonal.
Moreover, we have studied the correlations of the elements of the
neutrino mass matrix.
The result of this analysis is useful for finding 
textures in the neutrino mass matrix and, therefore, for 
model building in the lepton sector. As an example, we 
present two textures of the neutrino mass matrix which have
only two parameters and fit very well
all current experimental data on the neutrino parameters.}

\FullConference{The European Physical Society Conference on High Energy
Physics \\
                 18-24 July, 2013\\
                 Stockholm, Sweden}

\begin{document}

\section{Correlations of the elements of the neutrino mass
matrix}\label{correlation-section}
Assuming Majorana neutrinos, in the basis
where the charged-lepton mass matrix is diagonal, \textit{i.e.}
$\mathcal{M}_\ell=\mathrm{diag} \left( m_e,\, m_\mu,\, m_\tau \right)$,
the mass matrix of the three light neutrinos is given by
\begin{equation}\label{Mnu-UPMNS}
\mathcal{M}_\nu = U_\mathrm{PMNS}^\ast\, \mathrm{diag}
\left( m_1,\, m_2,\,m_3 \right) U_\mathrm{PMNS}^\dagger. 
\end{equation}
Therefore, in this basis one finds that the absolute values of the
elements of $\mathcal{M}_\nu$ depend on 
nine parameters, namely
\begin{equation}\label{parameters}
m_0,\, \Delta m_{21}^2,\, \Delta m_{31}^2,\, \theta_{12},\,
\theta_{23},\, \theta_{13},\, \delta,\, \rho,\, \sigma. 
\end{equation}
Here $m_0$ denotes the mass of the lightest neutrino, the strongest bound on which
comes from cosmology. In our analysis we used the upper bound
$m_0 \leq 0.3\,\mathrm{eV}$. The allowed ranges for the six oscillation parameters were taken from
the global fits of oscillation data of~\cite{forero,fogli}.
The phases $\rho$ and $\sigma$ denote the two experimentally unconstrained Majorana phases, which can
thus assume any value between $0$ and $2\pi$.

Using the experimental input discussed above one
can numerically derive restrictions on the absolute values 
$|(\mathcal{M}_\nu)_{ij}|$ of the
elements of the neutrino mass matrix~\cite{rodejohann,correlations}
and also compute \textit{correlations} between these elements.
We present two examples of correlation plots,\footnote{The light
Majorana neutrino mass matrix has six independent entries, thus
there are 15 correlations. Since there are two possible mass spectra, there is
a total of 30 plots.}
both based on the global fit results of Forero et
al.~\cite{forero}, assuming a normal spectrum. 
The boundaries of the corresponding $n\sigma$-regions are indicated as follows:
best fit: 
\begin{large}$\ast$\end{large}, 
\textcolor{red}{$1\sigma$: $\blacktriangle$}, 
\textcolor{blue}{$3\sigma$: $\bullet$}.
\begin{center}
\includegraphics[width=0.4\textwidth,angle=-90]{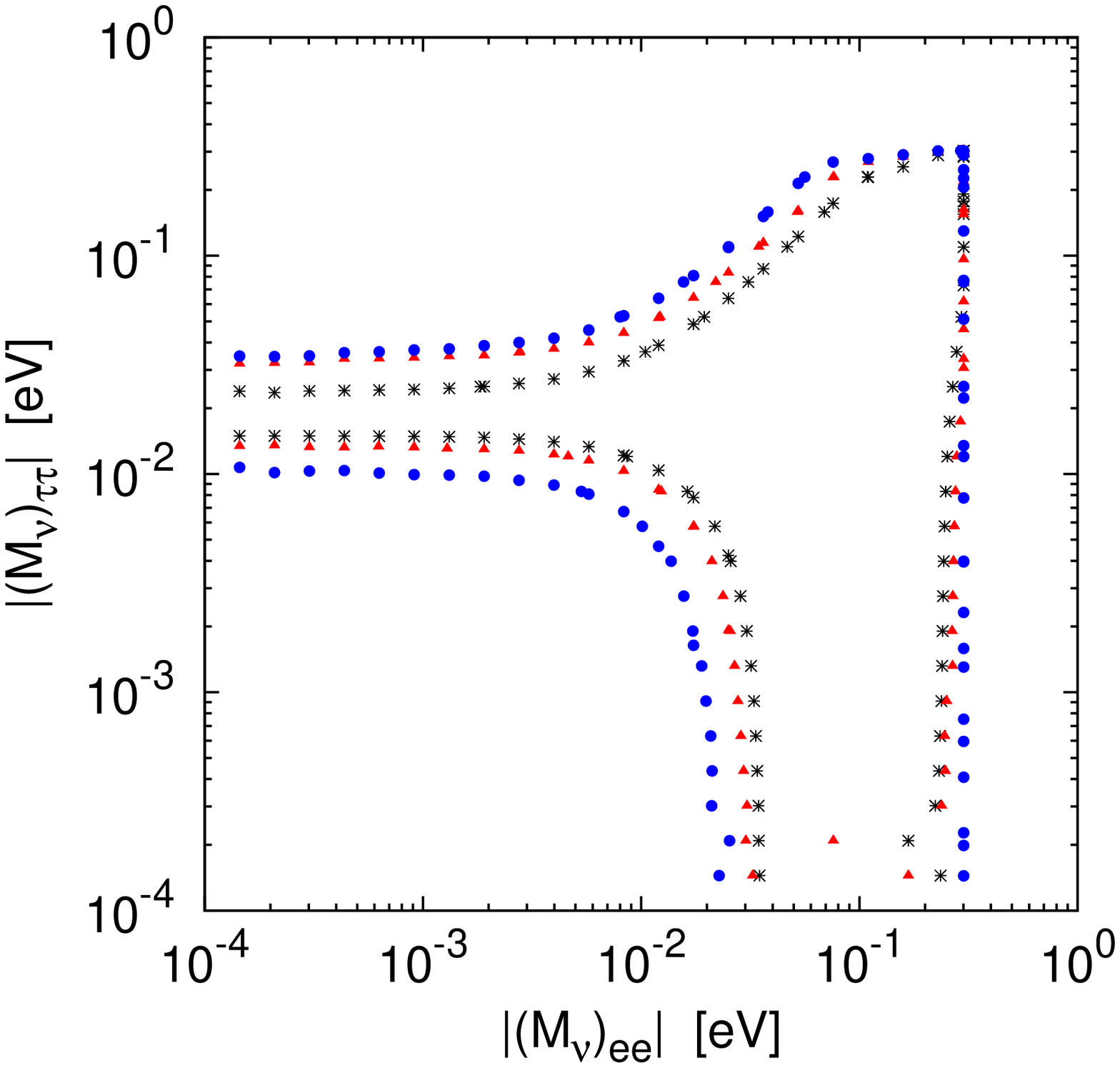}
\hspace{2cm}
\includegraphics[width=0.4\textwidth,angle=-90]{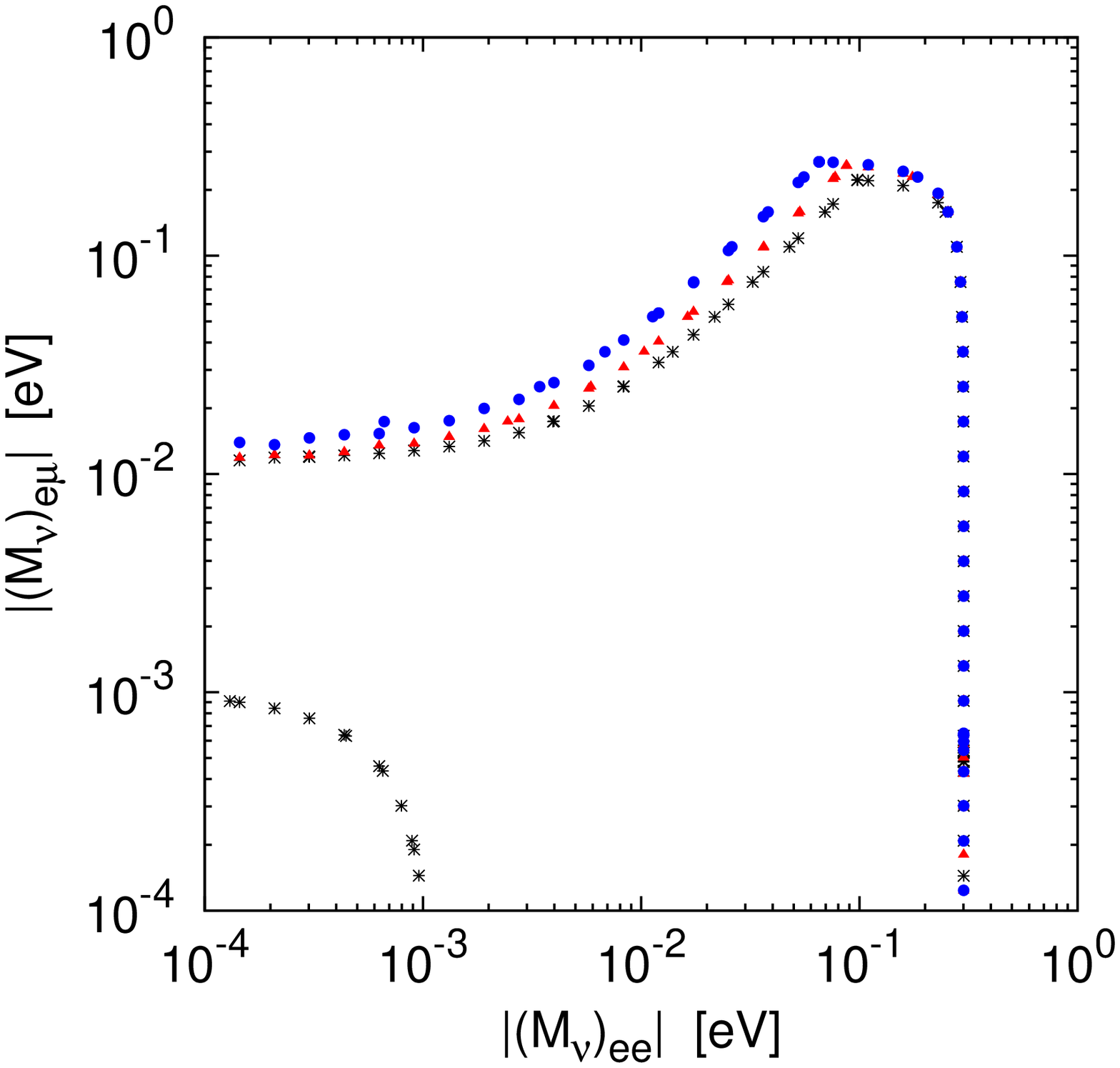}
\end{center}
From the first plot we see that, even at the 3$\sigma$-level, 
there is a strong correlation between
$|(\mathcal{M}_\nu)_{ee}|$ and $|(\mathcal{M}_\nu)_{\tau\tau}|$ which 
does not allow that the two matrix elements are small at the same time.
In fact, all correlations we found to be stringent at the 3$\sigma$-level
are of this type. The stringent correlations are the following ones:
\begin{center}
\begin{tabular}{llll}
$|(\mathcal{M}_\nu)_{ee}|$ & vs. & $|(\mathcal{M}_\nu)_{\mu\mu}|$ &
  (normal spectrum) \\
$|(\mathcal{M}_\nu)_{ee}|$ & vs. & $|(\mathcal{M}_\nu)_{\mu\tau}|$ &
  (normal spectrum) \\
$|(\mathcal{M}_\nu)_{ee}|$ & vs. & $|(\mathcal{M}_\nu)_{\tau\tau}|$ &
  (normal spectrum) \\
$|(\mathcal{M}_\nu)_{\mu\mu}|$ & vs. & $|(\mathcal{M}_\nu)_{\mu\tau}|$
  & (normal spectrum) \\
$|(\mathcal{M}_\nu)_{\mu\tau}|$ & vs. &
  $|(\mathcal{M}_\nu)_{\tau\tau}|$ & (normal spectrum).
\end{tabular}
\end{center}
For an inverted mass spectrum there are no correlations manifest at
the 3$\sigma$-level. 

Turning to the second plot we see that (in the case of a normal mass spectrum),
at 1, 2 and 3$\sigma$, $(\mathcal{M}_\nu)_{ee}$
and $(\mathcal{M}_\nu)_{e\mu}$
can simultaneously be very small, 
in contrast to the first plot.
This consideration leads to the concept of texture zeros in the
neutrino mass matrix~\cite{FGM}. Since they
are easily realizable by means of Abelian flavour
symmetries~\cite{Grimus-Joshipura}, texture zeros 
have become very popular in model building.
An obvious question arising
in this context is whether the neutrino mass matrix supports more
structure than just texture zeros, \textit{i.e.}, whether one can
impose additional relations leading to so-called 
``hybrid textures.''

\section{Hybrid textures of the neutrino mass matrix}

Our strategy to unveil possible ``hybrid textures'' of $\mathcal{M}_\nu$ is the following~\cite{hybrid}. We define
the function
\begin{equation}
F_{ijkl}\equiv |(\mathcal{M}_\nu)_{ij}| + |(\mathcal{M}_\nu)_{kl}|,
\end{equation}
which can be minimized within the allowed range of its nine parameters---see
equation~(\ref{parameters}).
A minimum of zero (within the numerical accuracy) displays an allowed case of
two texture zeros. This analysis does not only confirm the results
deduced from the 
correlation plots of section~\ref{correlation-section}, it also
provides an estimate of 
the neutrino mass matrix in a certain sense. Namely, taking the nine
parameters at the minimum of $F_{ijkl}$, we can estimate the values of
$|(\mathcal{M}_\nu)_{ij}|$ for all $i,j = e, \mu, \tau$. 
In this way we find two types of two texture zeros which can be endowed with a hybrid structure,
namely~\cite{hybrid}
\begin{equation}
\mathcal{M}_\nu = 
\left( \begin{array}{ccc}
0 & 0 & a \\ 
0 & b & 2a \\
a & 2a & b
\end{array}
\right)\quad\quad
\text{and}
\quad\quad \mathcal{M}_\nu = 
\left( \begin{array}{ccc}
0 & a & 0 \\ 
a & b & 2a \\
0 & 2a & b
\end{array}
\right).
\end{equation}
Surprisingly, the above textures are very well compatible with
observations even for real parameters $a$ and $b$. In this case,
the absolute neutrino mass scale, the three mixing angles and
the two mass-squared differences are functions of only two real parameters. 
\medskip
\\
\textbf{Acknowledgments:} This work is supported by the Austrian Science Fund (FWF),
Project No. P 24161-N16.

\end{document}